# Electronic self-doping of Mo-states in $A_2FeMoO_6$ (A=Ca, Sr and Ba) half-metallic ferromagnets – a Nuclear Magnetic Resonance study

M. Wojcik, E. Jedryka and S. Nadolski

*Institute of Physics, Polish Academy of Sciences, Al. Lotników 32/46, 02 668 Warszawa, Poland*

D. Rubi, C. Frontera and J. Fontcuberta,

*Institut de Ciència de Materials de Barcelona, CISC, Campus Universitari de Bellatera, E-08193 Bellatera, Catalunya, Spain.*

B. Jurca, N. Dragoe and P. Berthet

*Laboratoire de Physico-Chimie de l'Etat Solide, UMR-CNRS 8648, ICMMO, Bât.410, Université Paris Sud, F-91405 Orsay, Cedex, France*

**Abstract**

A systematic study of $(A,A')_2FeMoO_6$ (A,A'=Ca, Sr, Ba) ferromagnetic oxides with double perovskite structure has been performed using $^{95,97}Mo$ and $^{57}Fe$ NMR spectroscopy. These oxides are isoelectronic but have substantially different Curie temperatures. The NMR analysis provides clear evidence that the magnetic moment at Mo sites is not constant but varies sensitively with the ionic size of the alkaline ions. The $^{95,97}Mo$ frequency, and thus the electronic charge at Mo ions, is found to be smaller in Ba and Ca than in Sr-based oxides. The charge release from Mo sites is accompanied by an uptake at Fe sites, and thus a self-doping Fe-Mo process is observed. This process is controlled by relevant structural parameters: the Fe-O-Mo bond length and bending. A clear relationship between the Curie temperature and the magnetic moment and thus electron density at Mo sites has been disclosed. The relevance of these findings for the understanding of ferromagnetic coupling in double perovskites is discussed.

**Introduction**

Ferromagnetic coupling in half-metallic $L_{1-x}A_xMnO_3$ manganites is commonly described by the so called "double exchange model" [1]. Within this scenario, the electronic configuration of Mn ions rapidly fluctuates between $Mn^{3+}$ and $Mn^{4+}$ states and the Curie temperature ($T_C \leq 360K$) is determined by the hole concentration in the conduction band and its bandwidth (W). Double perovskites (DP), such as $A_2FeMoO_6$ (A= Sr, Ca, Ba), are also half-metallic ferromagnets that are receiving much attention due to the fact that their Curie temperature is even higher than that of manganites making them more promising materials for applications [2]. However, understanding the ferromagnetic coupling in these oxides is more challenging [3]: the experimental observation of a higher $T_C$ in DP sharply contrasts with the doubling of the separation between magnetic ions (basically the Fe ions). It has recently been proposed that $T_C$ in DP is directly related to the density of states at the Fermi level [4] and thus $T_C$ can be increased by electron doping [5]. Subsequent experiments have confirmed this prediction [6-10]. Moreover, it has been shown that indeed, the rise of $T_C$ is accompanied by an enhancement of the density of states at the Fermi level - mainly its projection onto Mo-based states [11]. In agreement with this finding, recent $^{95,97}Mo$ Nuclear Magnetic Resonance (NMR) experiments on electron doped $Sr_{2-x}La_xFeMoO_6$ have shown that there is a clear correlation between the Curie temperature and the magnetic moment at Mo ions [12].

Pristine $A_2FeMoO_6$ (A=Ca, Sr and Ba) oxides are formally isoelectronic, with a single electron occupying the *spin-down* subband formed by orbitals of $3t_{2g}(Fe)+4t_{2g}(Mo)+2p(O)$ parentages [2]. Interestingly enough, in spite of the fact that the filling of the conduction band is the same, these oxides have remarkably different Curie temperatures (~345-365K in $Ca_2FeMoO_6$ [6, 13, 14, 15], 400-425K in $Sr_2FeMoO_6$

[1, 5, 6, 14, 16] and 308-367K in $Ba_2FeMoO_6$ [13, 17], respectively), i.e. $T_C(Sr) > T_C(Ba)$, $T_C(Ca)$.

To account for this striking observation, and using a picture similar to that used in manganites, claims have been made that differences in $T_C$ are due to variations of the conduction bandwidth due to differences of the Fe-O-Mo bond topology, which are caused by the distinct size of the $A^{2+}$ cations [13, 18]. However, as already mentioned, it is not obvious that the double exchange picture can be used in this system. Moreover, while the bond bending in manganites is expected to modify the conduction bandwidth W without changing the charge density at any $Mn^{3/4+}$ ion, this is not the case in DP, where the presence of distinct Fe and Mo ions may promote substantially different results.

In this paper we shall address this issue by using $^{95,97}Mo$ and $^{57}Fe$ NMR as local probes to determine the magnetic moment (and thus electron density) at Mo and Fe sites in $A_2FeMoO_6$ (A=Ca, Sr, and Ba). We will show that the bond bending determines the $3d^6(Fe^{2+}) \leftrightarrow 4d^0(Mo^{6+})$ charge equilibrium and thus triggers a self-doping of Mo ions and a depletion of charge at Fe ions. The variation of $T_C$ in these compounds is found to be mainly related to the electron concentration at Mo sites. These findings, sharply contrasting with the physics of manganites, reveal the unique influence of the filling of Mo states on the strength of ferromagnetic coupling and shall provide a new insight into the nature of ferromagnetic state in these oxides.

**Experimental**

Two series of ceramic samples: $Sr_{2-x}Ca_xFeMoO_6$ (x=0, 0.4, 0.6 and 2.0) and $Sr_{2-x}Ba_xFeMoO_6$ (x=0, 0.2, 0.4, 1.0 and 2.0) have been synthesized as described in Ref. [6] and Ref. [17] respectively. X-ray diffraction and Rietveld refinement of the diffraction profiles have been used to refine the structure and to determine a fraction of misplaced Fe(Mo) ions at Mo(Fe) sites. These antisite defects (AS) are known to be common in DP and in case of present samples their content was: AS(Ca) = 6 %, AS(Sr) = 9 % and AS(Ba) = 5%. The structural data confirm that as the ionic radius of the alkalines increases from $Ca^{2+}$ (1.12Å) to $Sr^{2+}$ (1.26Å) to $Ba^{2+}$ (1.42Å), an expansion of the unit cell volume takes place [6, 13]. Magnetization measurements have been performed using a SQUID magnetometer. The saturation magnetization ($M_S$) and the Curie temperature of the Ca, Sr and Ba based samples, extracted from such measurements are: 3.7 $\mu_B$/f.u., 3.4 $\mu_B$/f.u. and 3.6 $\mu_B$ and 375 K, 397K and 308K, respectively.

NMR spectra have been recorded with a resolution of at least 0.5 MHz (0.25 MHz where necessary), in the frequency range between 20 MHz and 200 MHz, using an automated, coherent, phase sensitive spin-echo spectrometer [19]. Experiments have been carried out at 4.2 K in zero external magnetic field for different amplitudes of r.f. pulses. NMR signal intensity has been corrected for the intrinsic enhancement factor, which was experimentally determined at each frequency point from the spin echo intensity dependence on the excitation r.f. power level [20]. It must be stressed that the $^{95,97}$Mo NMR spectra in $Sr_2FeMoO_6$ (SFMO) were found to be totally insensitive to the AS content – a feature attributed to the half metallic character of these materials and the lack of s-type conduction electrons [12, 21].

**Results and discussion**

The NMR spectra recorded from the series of $(Sr_{2-x}Ca_x)FeMoO_6$ samples are presented in Fig.1. In agreement with previous reports [12,21,22], the NMR spectrum corresponding to the undiluted (x=0) SFMO sample presents a mixture of $^{95}$Mo and

$^{97}$Mo resonances (their giromagnetic ratio being too close to each other to separate the respective signals) while the weak $^{57}$Fe NMR line has an overlapping frequency (65.8 MHz) and is not resolved in this case. The $^{95,97}$Mo NMR spectrum (top panel) is dominated by the main resonance line at $\nu_{Mo}\sim 66.8$ MHz. The low frequency structure, extending down to about 40 MHz, is not fully understood although it has been suggested [12] that it could be a manifestation of charge instability [3, 23]. Upon dilution with Ca, the main Mo NMR line broadens up and for x=2 (CFMO compound) it is shifted down to 62.6 MHz. At the same time, the $^{57}$Fe NMR resonance becomes visible in form of a small shoulder at $\nu_{Fe}\sim 65.8$ MHz that can be resolved against the background of the Mo resonances in the NMR spectra of the x=0.4, 0.6 and x=2.0 samples. The distinct origin of this subtle spectral feature is additionally confirmed by its different restoring field [20]. Indeed, in the plot of the frequency dependence of the NMR restoring field, shown in the bottom panel of Fig. 1, the small peak at $\nu_{Fe} \approx 65.8$ MHz clearly displays a different response to the r.f. excitation. We thus attribute this feature to the $^{57}$Fe signal, in agreement with our previous NMR study [12] and with results of Mössbauer experiments [24, 25, 26]. We note in Fig. 1 that the position of this line ($\nu_{Fe}$) remains almost constant throughout the $Sr_{2-x}Ca_xFeMoO_6$ series of samples. We remind here that a similar behavior, i.e. a constant $^{57}$Fe NMR signal against the background of a shifting $^{95,97}$Mo spectrum has been reported for $(Sr_{2-x}La_x)FeMoO_6$ samples [12].

The NMR spectra recorded from a series of $(Sr_{2-x}Ba_x)FeMoO_6$ samples are presented in Fig.2. We note in the spectra that upon Ba substitution there is a clear downward shift of both Mo and Fe NMR frequencies. In the pristine $Ba_2FeMoO_6$ (BFMO) compound, the $^{57}$Fe NMR line can be clearly resolved at $\nu_{Fe}\sim 63$ MHz, as shown in the bottom panel in Fig.2, where we additionally plot a part of the spectrum in an expanded intensity scale, in order to bring out the details. $^{57}$Fe hyperfine field (45.8 T) determined from NMR frequency is in perfect agreement with Mössbauer results (46 T) [25]. The main $^{95,97}$Mo line in BFMO is also shifted downwards with respect to SFMO and its frequency position is 55 MHz.

To summarize and compare the respective NMR line positions for the two studied series of samples, in Fig.3 we juxtapose the spectra recorded from the three pristine compounds (SFMO, CFMO and BFMO). Fig. 4 collects the variation of resonance frequency ($\nu_{Mo}(x)$) of the main line ($^{95,97}$Mo NMR) (bottom panel) and the position of the $^{57}$Fe line ($\nu_{Fe}(x)$) (top panel) versus the (Ca, Ba) concentration x (bottom axis). Data in this figure already indicate the most remarkable finding, namely: upon expanding the unit cell volume from Ca to Sr to Ba [6, 13], the resonance frequencies of Mo and Fe nuclei vary in a non-monotonic way. More precisely, $\nu_{Mo}(x)$ has a maximum around the SFMO composition and decreases –non symmetrically– for BFMO and CFMO. Similarly, $\nu_{Fe}(x)$ decreases from SFMO to BFMO but remains almost unchanged from SFMO to CFMO.

As mentioned above, due to the lack of s-electrons in the conduction band of SFMO, the transferred field mechanism is inactive and the NMR frequency ($\nu$) is directly proportional to the on-site atomic magnetic moment ($\mu$) [12,21]:

$$\nu = \gamma\, A_{core}\, \mu \quad (1)$$

where $\gamma$ is the nuclear giromagnetic ratio ( $\gamma$ ($^{95,97}$Mo)=2.796MHz/kOe - taken as the average between $\gamma$ ($^{95}$Mo) = 2.774 (15.78%) and $\gamma$ ($^{97}$Mo) = 2.833 (9.6%) - and $\gamma$($^{57}$Fe)=1.38 MHz/kOe ) and $A_{core}$ is the hyperfine interaction constant describing the core electron contribution to the contact term of hyperfine field for the respective nuclei ($^{95}$Mo, $^{97}$Mo and $^{57}$Fe) [12]. Therefore, the frequency positions ($\nu_{Fe}(x)$ and $\nu_{Mo}(x)$) shown in Fig. 4 imply that the magnetic moment in the corresponding ions is changing with the increasing substitution. This observation sharply contrasts with the fact that (Ca, Ba) substitutions in SFMO are isolectronic and thus modifications of the (Fe, Mo) atomic moments were not, *a priori*, expected.

To understand this striking behaviour it is interesting to consider the structural modifications accompanying the (Ca,Ba) substitutions in SFMO. As a measure of structural changes we have used the average atomic radius $<r_A>$ (weighted cationic radius of ions at A position in $A_2FeMoO_6$). Indeed, it is known that for large $<r_A>$ ions, the unit cell symmetry tends to be cubic (the BFMO case), whereas for smaller $<r_A>$ the unit cell becomes tetragonal (the SFMO case) or monoclinic (CFMO case). Using the top axis in Fig. 4 we present the respective $\nu_{Fe}(x)$ and $\nu_{Mo}(x)$ NMR frequencies as a function of the $<r_A>$ values for the studied compounds. This presentation makes clear that, in spite of a rather monotonous cell expansion upon substitution [6,13], the NMR frequencies $\nu_{Fe}(x)$ and $\nu_{Mo}(x)$ (and thus the corresponding $\mu$) display an unexpected non-monotonic variation. The Mo NMR frequency has a maximum for SFMO ($\nu_{Mo} \approx$ 66.8MHz) and decreases by $\Delta\nu_{Mo} \approx$ -11.8 MHz when going from SFMO to BSFMO and lowers $\Delta\nu_{Mo} \approx$ -4.3 MHz when going from SFMO to CFMO, as seen in Fig.5 (left axis scale). It is also clear from Figs. 4 and 5 that the Fe resonance frequency is less sensitive to the ionic radius. Indeed $\nu_{Fe} \approx$ 65.8MHz for SFMO and $\nu_{Fe} \approx$ 63.0MHz for BFMO, thus $\Delta\nu_{Fe} \approx$ 2.8MHz. The position of the Fe line in CFMO is $\nu_{Fe} \approx$ 65.8 MHz and thus no significant shift with respect to SFMO is observed.

Based on the NMR relationship in these materials, $\nu = \gamma A_{core} \mu$, frequency values can be transformed into the values of the on-site magnetic moment, using the appropriate nuclear parameters: $\gamma (^{57}Fe)$=1.38 MHz/kOe, and $A(^{57}Fe) \approx$ 100 kOe/$\mu_B$ [27]. For instance, $\nu_{Fe} \approx$ 65.8MHz, as observed in SFMO, corresponds to $\mu_{Fe} \approx$ 4.78$\mu_B$. We note that this value would indicate an electronic configuration of about Fe-3$d^{5.22}$, which is in good agreement with the Mössbauer data [24, 25]. It corresponds to a partial filling of the *spin-down* subband of Fe ions. Within the context of the present paper, it is even more important to notice that the observed decrease of $\nu_{Fe}$ ($\Delta\nu_{Fe} \approx 2.8$ MHz) from SFMO to BFMO corresponds to $\Delta\mu_{Fe} \approx -0.2$ $\mu_B$. A reduction of the atomic moment in Fe implies that the electronic concentration ($q_{Fe}$) in the Fe *spin-down* subband increases. In a localized ionic picture this would indicate a charge increase of $\Delta q_{Fe} \approx +0.2e$, where e is the electron charge. At the same time, any possible charge modification at Fe ions, when going from SFMO to CFMO, remains within the experimental resolution and thus the corresponding $\Delta q_{Fe} \approx 0$. The values of $\Delta q_{Fe}(x)$ for all samples are plotted in Fig. 5 (top-right axis).

As evidenced in Figs. 4 (bottom panel) and 5 (middle panel), more dramatic and well pronounced are the changes in the resonant frequency (and thus magnetic moment) of the Mo nuclei; the evaluation of the corresponding charge transfer values $\Delta q_{Mo}(x)$ would be desirable. Unfortunately accurate values of $A_{cor}(^{95,97}Mo)$ are not available: different values between 200 and 400 kOe/$\mu_B$ have been reported in the literature [27, 28]. Therefore a reliable direct estimation of $\Delta\mu_{Mo}(x)$ from data in Fig. 4 is not possible.

To overcome this difficulty it is interesting to remind here that we have recently reported a clear upshift of $\nu_{Mo}(x)$ upon electron doping -achieved *via* trivalent lanthanide substitution- in SFMO (or CFMO) [12, 29]. In Fig. 6 we show the values of $\nu_{Mo}(x)$ (main line) obtained for $Sr_{2-x}La_xFeMoO_6$ (data taken from Ref.12.), $Ca_{2-x}Nd_xFeMoO_6$ (from Ref. 29), $Sr_{2-x}Nd_xFeMoO_6$ (from Ref. 30). It is clear from Fig. 6 that all data points converge into a single straight line indicating that the electron filling of the Mo orbitals determines the NMR frequency and thus other possible structural modifications associated with the atomic substitution should have a minor impact on $\nu_{Mo}(x)$. Consequently, we can safely assume that the fitted slope $d\nu_{Mo}/dx=57.6$

MHz/electron from the data in Fig. 6, corresponds mainly to the electronic filling effect and thus this slope provides an accurate estimate of the effectiveness of carrier doping on the Mo frequency shift. It follows that this $d\nu_{Mo}/dx$ value (~57.6 MHz/electron) can be used to evaluate the $\Delta q_{Mo}(x)$ from the $\Delta\nu_{Mo}(x)$ data of Fig. 5. For instance: when going from SFMO to BFMO, $\Delta\nu_{Mo}(x)\approx-11.8$MHz what corresponds to $\Delta q_{Mo}(x) \approx -0.21$e. Thus in BFMO the Mo-4d spin down states are charge depleted by some -0.21e. Similarly, for CFMO $\Delta q_{Mo}(x) \approx -0.08$e. In Fig, 5 (middle panel) we show the charge variation at Mo sites ($\Delta q_{Mo}(x)$) evaluated in this way, as a function of x for all compounds studied.

    The internal consistency of this analysis is rewarding. Comparing the data in top and middle panels in Fig. 5 it is clear that charge released from Mo states when going from SFMO to BFMO ($\Delta q_{Mo}(x) \approx -0.21$e) is transferred to Fe states. Moreover, $\Delta q_{Mo}(x) \approx -0.21$e as determined for BFMO compares very well with the charge uptake by Fe as deduced from the shift of the Fe line (~0.2e). The Mo states in $Ca_2FeMoO_6$ are also depleted when compared to those of $Sr_2FeMoO_6$ although a smaller fraction of charge is transferred ($\Delta q_{Mo}(x) \approx -0.08$e). If charge were transferred solely to Fe states, then a corresponding uptake of charge by Fe and thus a frequency decrease $\Delta\nu_{Fe}$ of about 1 MHz should occur. As revealed by data in Fig. 5 (top panel), within the experimental resolution this shift of $\Delta\nu_{Fe}\approx 1$MHz is not observed as $\nu_{Fe}(SFMO) \approx \nu_{Fe}(CFMO)$. The invariance of the Fe magnetic moment in SFMO and CFMO is in agreement with the observation of very similar isomer shift values in the corresponding Mössbauer spectra [25]. The fact that the Mo states are modified whereas Fe states remain unaltered, suggest that in CFMO, the oxygen 2p-orbitals largely contribute to the conduction band. The self-doping mechanism we have described is schematically depicted in Fig. 7. In

short, we conclude that the integrated 4d-charge density, and thus the magnetic moment of Mo ions in CFMO and BFMO is reduced with respect to SFMO.

The relevance of these observations becomes clear when the Curie temperatures of these isoelectronic series of compounds are plotted (see Fig. 5 (bottom panel)) and compared with the magnetic moment of Mo ions ($\mu_{Mo}$). It is obvious that there is a direct relationship between $\mu_{Mo}$ and $T_C$. This can be fully appreciated in Fig. 8 where we plot the Curie temperatures versus the Mo resonance frequency for all $A_{2-x}A_x'FeMoO_6$ (A,A'= Ca, Sr and Ba) samples reported in this study. The $T_C(\nu_{Mo})$ for all samples roughly fall on a single straight line ($dT_C/\nu_{Mo} \approx 7.4$ K/MHz), suggesting indeed that the magnetic moment at Mo sites largely determines the Curie temperature of these compounds.

On the basis of a simple structural picture, we could expect that electronic bandwidth W of pristine compounds varies as W(Ca)<W(Sr) and W(Ba)<W(Sr). This non-monotonic variation of W with the ionic size of the A cations follows from the fact that for small A cations (such as Ca), the structure is monoclinic and there is a substantial Mo-O-Fe bond bending that reduces W. On the other hand, for larger A cations (such as Ba), the structure is cubic, the Mo-O-Fe bond has an optimal overlapping (180º) but the unit cell expands do the size of the Ba ion, thus reducing again W; therefore, W should have a maximum somewhere around the $Sr_2FeMoO_6$ composition.

From the analysis of neutron diffraction structural data, Ritter et al [13] attempted to estimate the variation of the bandwidth W from the variation of bond angle ($\Theta$) and bond lengths ($d$). To that purpose an expression (W~ $\cos\Theta/d^{7/2}$) well suited for bonds of $e_g$ symmetry and a unique metal-oxygen bond-length was used [31]. The dependence of W on ionic size was found to mimic the observed $T_C$ dependence (shown in Fig. 5(bottom panel) and thus it was proposed [13] that W is the key parameter controlling the strength of the ferromagnetic coupling and thus $T_C$. However, it is unclear the accuracy of this approach to describe the bandwidth in the case of SFMO where relevant orbitals have a $t_{2g}$ symmetry and two-different bond lengths (Fe-O, Mo-O) are involved. More important, this picture, which is based on a model of double-exchange constructed for manganites, cannot be of direct application in DP since the electronic charge is redistributed among metallic ions as W is modified in DP, as we have shown in the present study.

At this point it is instructive to perform a detailed comparison of SFMO based double perovskites with ferromagnetic and metallic manganites. In $L_{1-x}A_xMnO_3$ (L=La, Nd,.. and A=Ca, Sr, Ba), at fixed charge doping (i.e x=1/3), the Curie temperature is determined by the bandwidth W of the conduction band [1]. When W broadens, for instance when going from $La_{2/3}Ca_{1/3}MnO_3$ to $La_{2/3}Sr_{1/3}MnO_3$, the Curie temperature is found to increase but there is no evidence of charge redistribution among metallic ions. Indeed, $^{55}$Mn-NMR results indicate a single resonant line centered at about ~377MHz which is essentially invariant upon W variation [32]. This implies that when W is varied the hole density is equally shared by all Mn ions in the structure. This is clearly in contrast to what we have observed in DP, where a self-doping process takes place.

It thus follows that electronic models aiming to get a microscopic understanding of the origin of the Curie temperature variation in the pristine compounds (CFMO, SFMO and BFMO) would require using precise structural data for each compound. Recently, Z. Szotek et al. [33] have calculated the electronic structure in the ferromagnetic phase of $A_2FeMoO_6$ (A=Ca, Sr, and Ba) and predicted that the magnetic moment at Mo sites in CFMO is somewhat smaller than that evaluated in SFMO. This estimate is in agreement with our observation.

Before closing we would like to address the origin of the charge redistribution which we have observed along these series of compounds. One can easily argue that when Ba is introduced in the structure the volume available for Mo and Fe ions expands in a significant way. In fact, it can be shown that when the substitution of Sr by Ba overpasses the tetragonal to cubic transition, the space available for the Fe/Mo ions becomes too large. Indeed, the increase of $<r_A>$ length can be compensated in the tetragonal structure by an increase of Mo-O-Fe bond angle, without a significant change in the Fe-O and Mo-O bond distances. When this bond angle reaches its maximum value (180°), the structure becomes cubic and any further enhancement of $<r_A>$ will lead to an expansion of the space available for Fe and Mo ions. In fact, from the bond distances reported in Ref. [6] the mean (Fe,Mo)-O bond distance in $Sr_2FeMoO_6$ is 1.974(2) Å. In the cubic structure, this mean bond-distance is constrained to be *a*/4, where *a* is the cell parameter of the cubic unit cell. For the particular case of $Ba_2FeMoO_6$ this is about 2.015 Å [13], a value considerably larger than the one corresponding to $Sr_2FeMoO_6$. It is well known that bond distances and valences are directly related, so a modification of the bond distance will drive a modification of the cationic valences. As given in Ref. [34] the effective ionic sizes of $Mo^{5+}$ and $Mo^{6+}$ are 0.61 Å and 0.59 Å respectively, and those of $Fe^{3+}$ and $Fe^{2+}$ are 0.645 Å and 0.780 Å respectively. From these values, for a larger A cation in $A_2FeMoO_6$ (such as in $Ba_2FeMoO_6$) the charge distribution will tend to be $Fe^{2+}/Mo^{6+}$ rather than $Fe^{3+}/Mo^{5+}$ as the corresponding (Fe,Mo)-O bond length will tend to be longer. In short, charge disproportionation results from the need of the electron structure to adapt the bond length to the interatomic distance imposed by the large Ba ion.

**Summary**

In summary, we have reported a systematic study of $^{95,97}$Mo and $^{57}$Fe NMR on two series of (A,A')$_2$FeMoO$_6$ (A,A'= Sr,Ca and Sr,Ba) samples. These oxides are isoelectronic and ferromagnetic but have substantially different Curie temperatures. The NMR analysis has clearly evidenced that the magnetic moment at Mo sites is not constant but can be tuned by the ionic size of the alkaline ions. The magnetic moment of molybdenum was found to be smaller in Ba and Ca than in Sr-based oxides. It has been shown that the charge release from Mo sites is accompanied by an uptake at Fe sites, and thus a self-doping Fe-Mo process takes place. This process is controlled by relevant structural parameters: the Fe-O-Mo bond length and bending, which in turn are determined by the size of the alkaline ion. Moreover, we have disclosed a clear relationship between the Curie temperature and the magnetic moment and thus the electron density at Mo sites. This finding should help to get a microscopic understanding of ferromagnetic coupling in double perovskites. From a more general perspective, our results and the use made of NMR spectroscopy illustrates the potentialities of this technique to monitor tiny changes of electron density in oxides. This can be useful in other active areas of research –such as ferroelectric materials.

**Acknowledgements**

This work has been supported in part by Research Framework Programme V (Growth) of the European Community under contract number G5RD-CT2000-00138" (AMORE), by a grant from Ford Motor Company (Poland), and by the MCyT (Spain) projects MAT2002-03431, MAT 2003-07483 and FEDER. C. F. acknowledges financial support from MCyT

**Figure Captions**

**Fig. 1** Zero field NMR spectra, recorded at 4.2 K from the series of $Sr_{2-x}Ca_xFeMoO_6$ of double perovskites. Predominant contribution to spectrum intensity arises from $^{95}Mo$ and $^{97}Mo$ nuclei. The spectrum feature at 65.8 MHz indicates $^{57}Fe$ NMR line position, which is additionaly confirmed by the plot of frequency dependence of the NMR restoring field (bottom panel). The contribution of $^{57}Fe$ NMR is much weaker due to the very low natural aboundance, smaller gyromagnetic ratio and lower nuclear spin.

**Fig. 2** Zero field NMR spectra recorded at 4.2 K from a series of $Sr_{2-x}Ba_xFeMoO_6$ double perovskites . A part of NMR spectrum for x=2 (bottom panel) is plotted in the enhanced intensity scale to resolve the $^{57}Fe$ NMR line at 63 MHz (see comments in the text). Upon increasing Ba content the frequency downshift of both, $^{95,97}Mo$ and $^{57}Fe$ NMR spectra is clearly evidenced.

**Fig. 3** Zero field NMR spectra recorded at 4.2 K from the pristine compounds $Ca_2FeMoO_6$, $Sr_2FeMoO_6$ and $Ba_2FeMoO_6$. The frequency position of $^{57}Fe$ NMR and main $^{95,97}Mo$ NMR are indicated.

**Fig. 4** The frequency position of the main $^{95,97}Mo$ and $^{57}Fe$ NMR lines as a function of doping concentration (bottom axis) and the mean size of A-site ionic radii ($<r_A>$, top axis), for $Sr_{2-x}Ca_xFeMoO_6$ and $Sr_{2-x}Ba_xFeMoO_6$ series of compounds.

**Fig. 5** Left axis : the NMR frequency shift with respect to the undiluted SFMO for $^{57}$Fe (top panel) and $^{95,97}$Mo (middle panel) versus the mean ionic radius of A-site ion. Right axis: the corresponding charge difference evaluated as described in the text. Bottom panel presents Curie temperature variation as a function of ionic radius.

**Fig. 6.** The ($^{95,97}$Mo) NMR frequency measured in the series of electron doped oxides $Sr_{2-x}La_xFeMoO_6$ (from Ref. 12), $Ca_{2-x}Nd_xFeMoO_6$ (from Ref. 29) and $Sr_{2-x}Nd_xFeMoO_6$ (from Ref. 30).

**Fig. 7** A schematic illustration of self-doping process involving transfer of charge $\Delta q$ between Mo 4d ($t_{2g}$) and Fe 3d($t_{2g}$) states. Spin down electron transfer $\Delta q$ between Mo and Fe reduces both, Fe and Mo magnetic moments and corresponding NMR frequencies.

**Fig. 8** The Curie temperatures of all compounds of the series $(Sr_{2-x}Ca_x)FeMoO_6$ and $(Sr_{2-x}Ba_x)FeMoO_6$ versus the corresponding $^{95,97}$Mo NMR frequency.

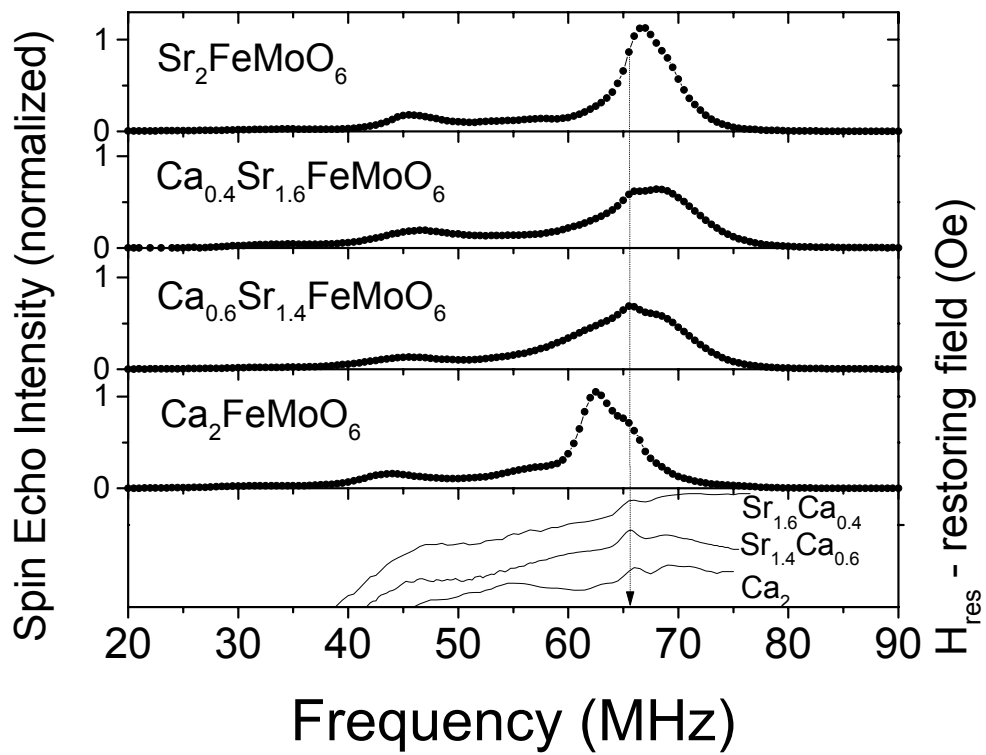

**Figure 1**

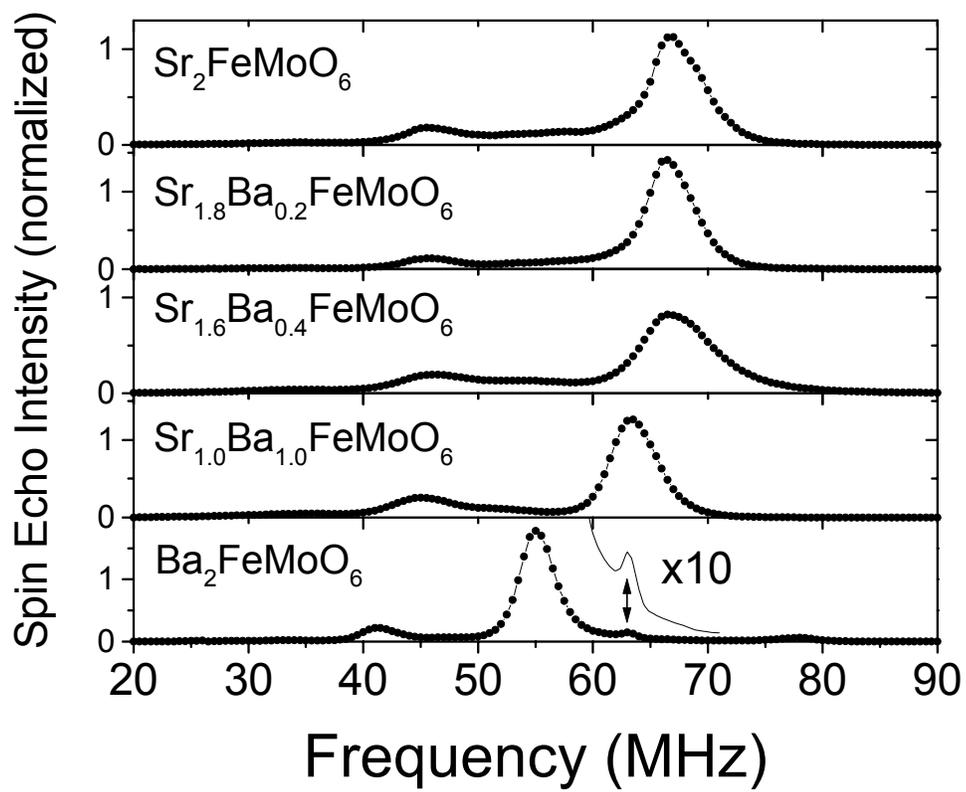

**Figure 2**

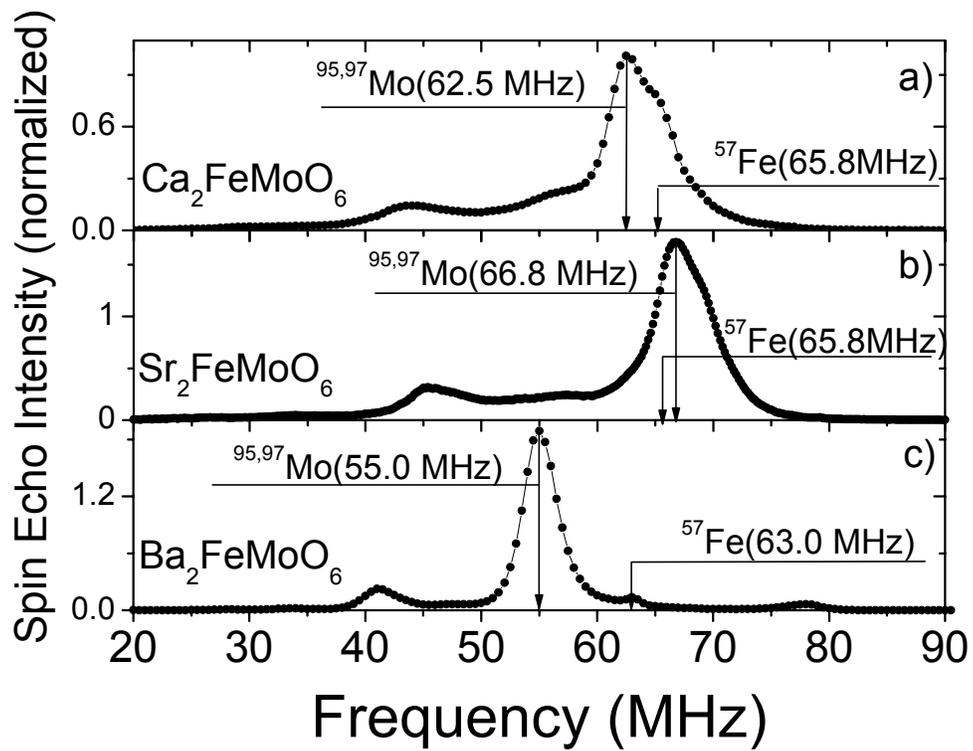

**Figure 3**

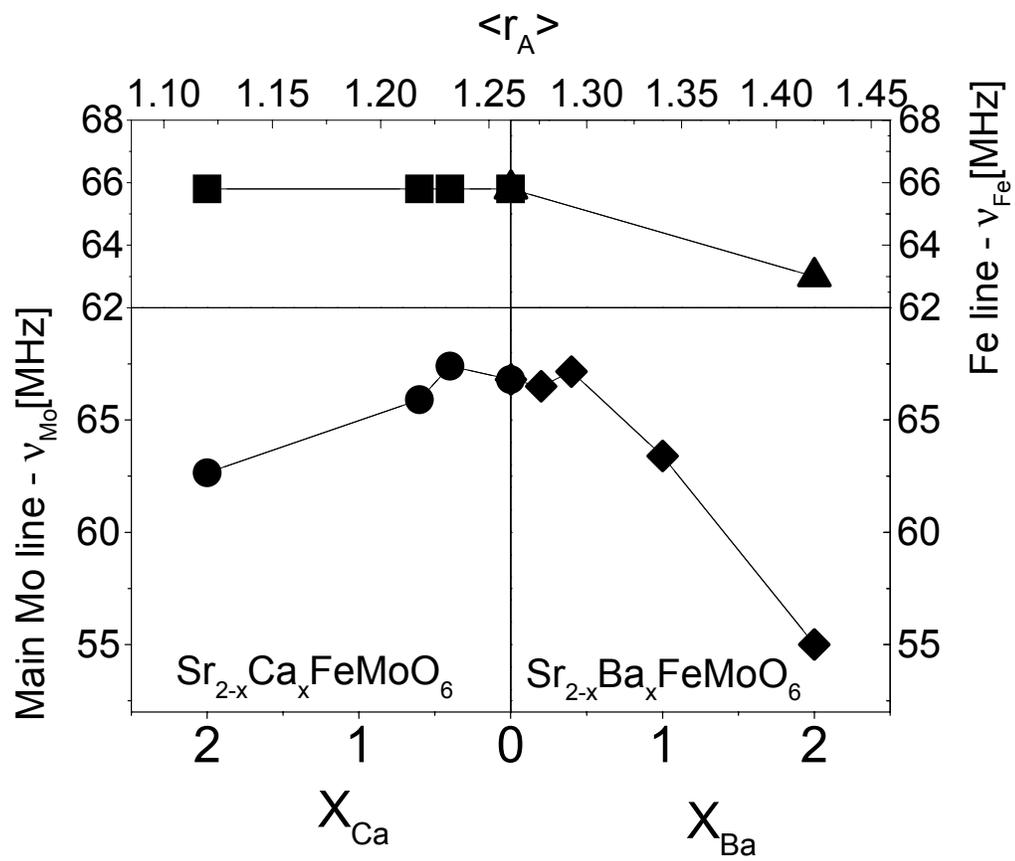

**Figure 4**

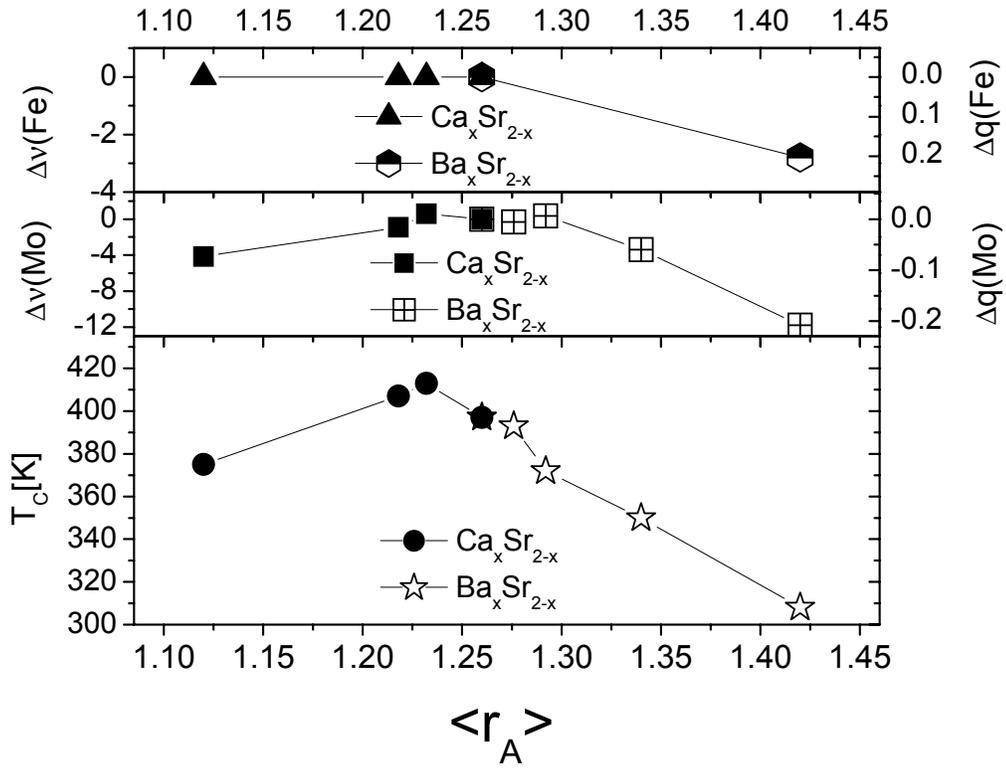

**Figure 5**

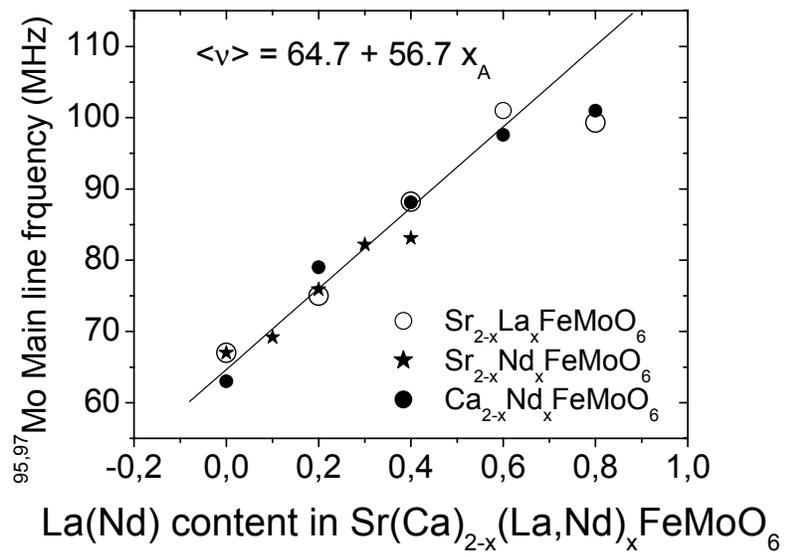

**Figure 6**

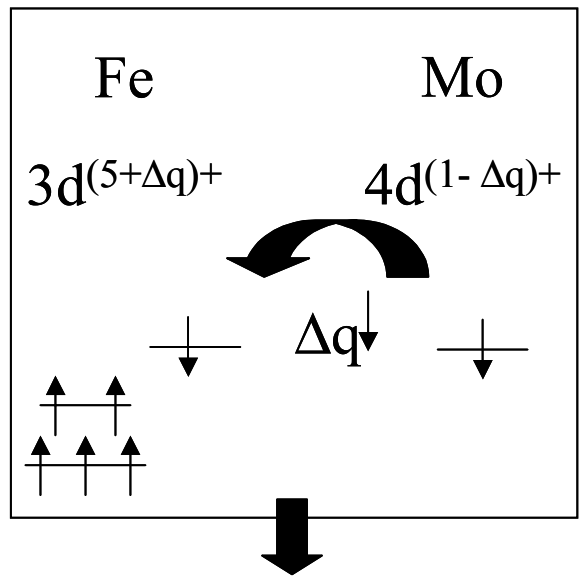

HF(Fe) decreases

HF(Mo) decreases

**Figure 7**

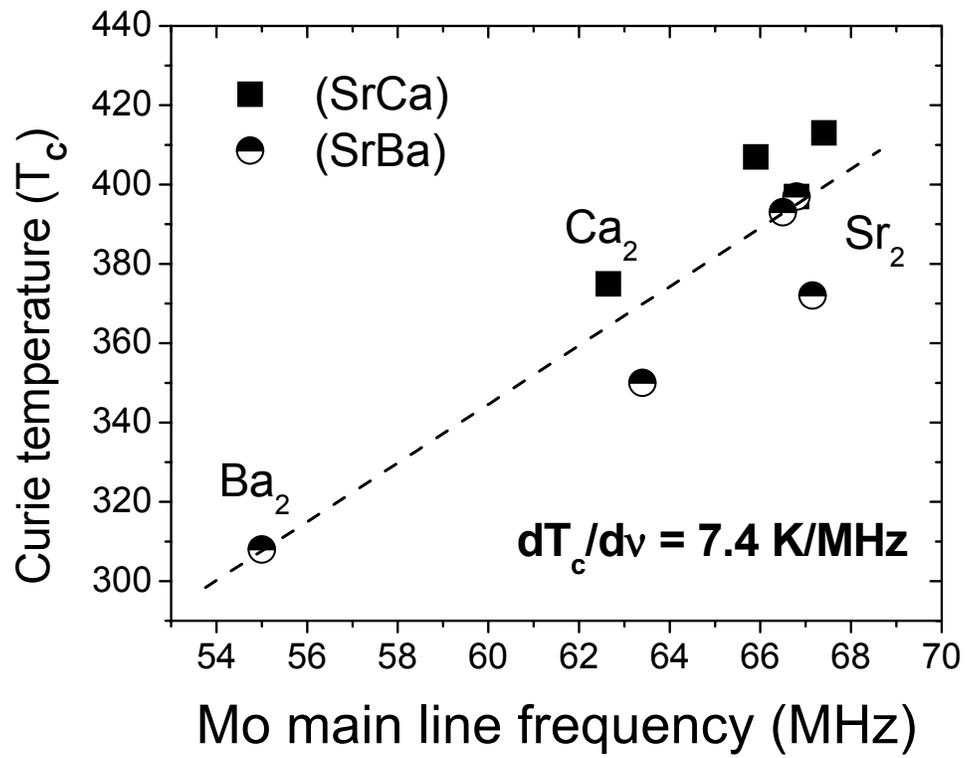

**Figure 8**